\documentclass[iop]{emulateapj}
\usepackage[]{graphicx}
\usepackage{amsmath}
\begin{document}
\setcounter{dbltopnumber}{1}
\title{A High-Entropy Wind R-Process Study Based On Nuclear-Structure
Quantities From The New Finite-Range Droplet Model Frdm(2012)}
\author{Karl-Ludwig Kratz\altaffilmark{1,2}} 
\author{Khalil Farouqi\altaffilmark{3,4}}
\author{Peter M{\"{o}}ller\altaffilmark{5}}
\affil{\altaffilmark{1}Max-Planck-Institut f\"{u}r 
Chemie (Otto-Hahn-Institut),
D-55128 Mainz, Germany;klk@uni-mainz.de}
\affil{\altaffilmark{2}Fachbereich Chemie, Pharmazie und 
Geowissenschaften, Universit{\"{a}}t Mainz,
         D-55128 Mainz, Germany}
\affil{\altaffilmark{3}
Zentrum f\"{u}r Astronomie der Universit\"{a}t Heidelberg, Landessternwarte,
D-69117 Heidelberg, Germany;kfarouqi@lsw.uni-heidelberg.de}
\affil{\altaffilmark{4}
Department of Physics, University of Basel, Klingelbergstrasse 82, CH-4056 Basel, Switzerland}
\affil{\altaffilmark{5}Theoretical Division MS B214, Los Alamos National Laboratory,
Los Alamos, NM 87545, USA;moller@lanl.gov}
\begin{abstract}
Theoretical studies of the nucleosynthesis origin of the heavy
elements in our Solar System (S.S.) by the rapid neutron-capture process
(r-process) still face the entwined uncertainties in the possible
astrophysical scenarios and the nuclear-physics properties far from
stability. In this paper we present results from the investigation of
an r-process in the high-entropy wind (HEW) of core-collapse
supernovae (here chosen as one of the possible scenarios for this nucleosynthesis process), using new nuclear-data input calculated
in a consistent approach, for
masses and $\beta$-decay properties from the new
finite-range droplet model FRDM(2012). 
The accuracy of the new mass model is 0.56 MeV with respect to {\sc AME2003}, to which it was adjusted. 
We compare the new HEW r-process abundance pattern to the latest
S.S. r-process residuals and to
our earlier calculations with the nuclear-structure quantities based
on FRDM(1992). Substantial overall and specific local improvements in
the calculated pattern of the r-process 
between $A\simeq 110$ and $^{209}$Bi, as well as
remaining deficiencies are discussed in terms of the underlying
spherical and deformed shell structure far from stability.
\end{abstract}

\section{INTRODUCTION}
Among the most challenging tasks in nuclear astrophysics is the theoretical description of a rapid neutron-capture nucleosynthesis process (the historical r-process) which is considered to be the origin of about half of the heavy element nuclides beyond Fe in our Solar System. This long-standing problem has been considered
number three among: ``{\it The 11 Greatest Unanswered Questions in Physics}'' \citep{haseltine02}.

Since the seminal papers of \citet{burbidge57,cameron57} and \citet{coryell61},
who identified the basic astrophysical and nuclear
conditions for an r-process in explosive environments with high neutron
density and temperature, r-process research has been quite diverse in
terms of suggested stellar scenarios (for representative reviews, see
e.g. \citep{cowan91,thielemann11}).
Still today, a robust production site for a full r-process with 
Solar-System-like
abundances up to the actinide chronometers $^{232}$Th and $^{238}$U has yet to be determined.

To experimental and theoretical nuclear physicists and chemists, modeling
the r-process has remained a particularly fascinating challenge over more
than 50 years. Its detailed study in terms of isotopic and elemental abundances
requires input of nuclear-structure data from the line of $\beta$-stability to the
neutron drip line. Simple parameterized, but nevertheless informative studies
of the r-process have been presented by many authors based on input of
just a few nuclear properties, namely the nuclear mass (from which neutron
separation energies $S_{\rm n}$ and $\beta$-decay $Q_\beta$ values can easily be
calculated), the $\beta$-decay half-lives $T_{1/2}$ and the probabilities of $\beta$-delayed
neutron emission $P_{\rm n}$. More elaborate, dynamical studies require additional
nuclear quantities, such as reaction rates, fission barriers, fission-fragment
yields and temperature dependencies of various parameters.

A considerable leap forward in the basic understanding of the required
astrophysical conditions at the time of r-process freeze-out for neutron
densities  occurred  about two decades ago \citep{kratz93}. At that time, nuclear data from earlier versions of our global,
unified, microscopic nuclear-structure models for nuclear masses (FRDM(1992)), published in \citet{moller95} 
and $\beta$-decay (the deformed quasi-particle random-phase (QRPA) model for Gamow-Teller (GT) $\beta$-decay
in the form of  \citet{moller97}, based on the original work
by \citet{krumlinde84}), together with the first few experimental data on r-process isotopes at $N=50$ and $N=82$ (see, e.g. ~\citet{kratz88}), were used for the first time in such calculations.
Among the main results of this investigation was a more detailed understanding of the impact of shell structure far from stability, considerably beyond the early recognition of the importance of spherical neutron shell closures. 
For example, the occurrence of the big r-abundance trough in the transitional mass region 115 $\le A \le 125$ was explained, for the first time the astrophysical consequences of a significant weakening of the $N = 82$ shell strength below doubly magic $^{132}$Sn were predicted, and evidence for a dramatic lowering of the neutron-g$_{7/2}$ orbital was found~\citep{kratz93}. Motivated by these findings we initiated a large series of nuclear-structure experiments at CERN/ISOLDE, which in the following years led to the identification of more than 20 new r-process-relevant isotopes in the $A \simeq 130$ mass region (see, e.g.~\citet{kautzsch2000,hannawald2000,dillmann2002,dillmann2003,shergur2002,shergur2005,kratz2005,arndt11}), as well as the collection of all relevant neutron and proton-particle and hole states  around $^{132}$Sn (see, e.g.~\citet{kratz2000}), including first evidence of the most difficult to identify p$_{1/2}$ and p$_{3/2}$ proton-hole states and their spin-orbit splitting in $N = 82$ $^{131}$In \citep{arndt2009}.

Also concerning the FRDM(1992) mass model, over the next several years, step by step improvements of the FRDM mass model have been implemented but only now finalized. Similarly, numerous enhancements have been added to the initial QRPA model, and an extensive discussion of the improved version has been published in \citet{moller03}. We therefore have available new consistent sets of the most essential nuclear-physics data.

With respect to our r-process parameter studies, since 1993 the
site-independent waiting-point approach was steadily refined
over the subsequent decade and applied to various comparisons with classical
and new astronomical observations~\citep{cowan99,pfeiffer01,sneden03,kratz07,ott08}.
Starting in the early new millennium, we  have 
replaced the above  approach with more realistic, site-dependent
dynamical r-process calculations within the high-entropy wind (HEW)
scenario of core-collapse supernovae (cc-SN).  In this effort,  the basic
Basel model~\citep{freiburghaus99} has been extended and
improved by \citet{farouqi05,farouqi09a,farouqi10}, 
and is still being used by
our group today. With this HEW model, several important
parameter studies have been performed to explain various
recent astronomical observations (see, e.g. \citet{farouqi09a,farouqi10,roederer09,roederer10,hansen12}). 

The aim of the present Article is to present dynamical nucleosynthesis
calculations within the HEW scenario, with the use of a new, improved
global set of nuclear-physics quantities based on the new
mass model FRDM(2012) and deformed QRPA calculations of $\beta$-decay properties.
We point out the major improvements in the
overall reproduction of the S.S. isotopic r-abundance residuals \citep{lodders09,bisterzo11}, as well as some remaining local deficiencies, which are smaller
than previously but can  provide important insights.
Our FRDM and QRPA model combination provides not only masses and half-lives for
the r-process calculation itself but many other associated quantities, e.g.
deformations. This in-depth information permits us to propose more 
definite conclusions about nuclear structure far from stability
through  detailed comparisons between calculated results and  observations.

\section{MODELS}

\subsection{The FRDM(2012) mass \label{2p1}model}

The first Los Alamos macroscopic-microscopic mass model, based
on the modern folded-Yukawa single-particle potential with
a globally optimized \citet{moller74} spin-orbit force was
published in 1981 \citet{moller81}. An enhanced version,
the finite-range droplet model, FRDM(1992),
which also extended the region considered to include all nuclei
between the proton and neutron drip lines,
was published in 1995
\citep{moller95}. This version now has been further refined during
the first 13 years of the new millennium in 6 steps, 5 of which are enumerated
in Figure 1 in \citet{moller12}. The final
improvement, leading to the new mass table FRDM(2012) is a more accurate
calculation of zero-point energies which further decreases the model error by
0.01 MeV, and removes well-known discontinuities in $S_{\rm 1n}$.
With the completion of these
steps, we  froze the model on 09/06/2012,  and are working
on submitting FRDM(2012) to {\sc Atomic Data and Nuclear Data Tables}.
The accuracy with respect to AME2003 is 0.56 MeV, but higher, 0.42 MeV,
above $N \simeq 50$, the region relevant for the r-process.

It is important to recall again that in our macroscopic-microscopic
approach a mass table is not only a set of nuclear masses (or total
binding energies), but just one result of a larger
theoretical and computational effort that also
provides ground-state (g.s.) shapes,
g.s. spins and parities, gross $\beta$-decay properties, 
decay spectra in terms of allowed GT
transition log ft values to daughter states, and many other related
quantities. There is also an associated single-particle-model computer
code so that one can with straightforward development obtain matrix
elements of other operators of interest, for example electric
quadrupole moments and charge radii.  For our current r-process studies, we have
calculated new tables of $T_{1/2}$ and $P_{\rm {\nu}n}$
based on $Q$-values and ground-state deformations from the new mass table.
 Deformations determined from
detailed calculations of multi-dimensional potential-energy surfaces
are essential for obtaining
realistic  $\beta$-decay transition spectra  from the parent ground states as
well as from excited states of possible $\beta$-isomers.
Our work here is unique in the sense that it obtains all
nuclear data needed  (ground-state deformations,
masses, half-lives and delayed-neutron
emission probabilities) from a single, consistent, global model framework,
which has been favorably compared to a substantial body of experimental data.
Furthermore, we obtain quantities  not just for even-even nuclides, as do
for example \citet{stoitsov03,delaroche10,niu13} but also for odd-even and odd-odd
nuclides, with level structure and the associated decay
schemes based on realistic intrinsic deformations.
This allows us to interpret the inevitable remaining differences
between calculations and observational data in terms of physics not yet
included in the model, or, possibly as issues with the experimental
evaluations. Excellent examples of previous such developments are,
for example, 1) in theory:
large differences between masses calculated in the original 1981
mass model \citet{moller81} and measured
masses  in the Ra region were removed by inclusion of octupole-type
shape distortions in the model deformation
space \citet{moller81,moller06},
and 2) in experiment: masses that were in the AME1993 data base
that deviated strongly from our calculations were removed 
or reevaluated in
the AME2003 data base, see  \citet{moller07} for a discussion.

\begin{figure}[t]
 \begin{center}
 \includegraphics[width=\linewidth]{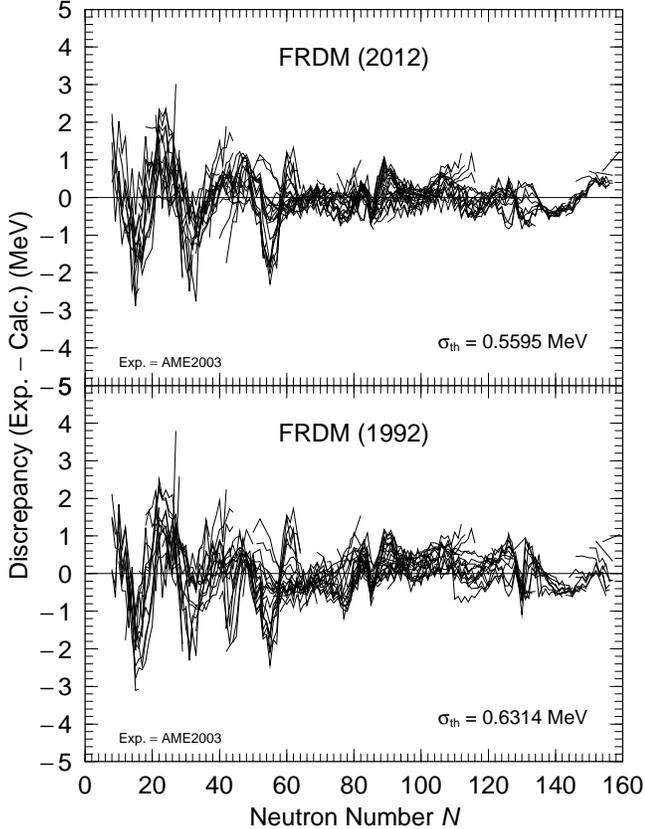}
   \caption{The FRDM(1992) and the  FRDM(2012) both compared
to the same experimental data set AME2003.}
\label{95vs12}
 \end{center}
\end{figure}

We show in Figure \ref{95vs12} the differences between calculated and experimental
masses from the 2003 mass evaluation of \citet{audi03} in our current
FRDM(2012) and the previous FRDM(1992).
At first sight, the new model
may not seem significantly better than the old one, but this is a fallacy.
The overall model error has been improved from $\sigma_{\rm th} = 0.6314$ MeV
to $\sigma_{\rm th} = 0.5595$ MeV, that is by about 13\%. Although the region
below $N \simeq 66$, that is below neutrons mid-shell
between $N = 50$ and $N=82$ shows larger correlated, fluctuating
deviations near closed shells and in the
transitional region around the sub-shell $N = 56$ compared
to the heavier region,
these deviations have minimal impact on
our HEW r-process calculations. In the heavier
region above,  the accuracy is significantly improved with
$\sigma_{\rm th}=0.36$ MeV. Detailed areas of improvements are:
1) the bump of negative deviations just beyond $N=40$ has largely
disappeared, 
2) the negative bump at $N \simeq 76$ is reduced, due to the
incorporation of axial asymmetry,
3) the large-amplitude fluctuations in the deviations present
in the vicinity of $N = 82$ are much reduced, especially below
the magic number, and, as a final observation
4) the large fluctuations near $N = 126$ have almost disappeared.

Moreover, other very short-range fluctuations in the deviations
are reduced. These fluctuations give large errors in calculated mass derivatives,
such as isobaric mass differences ($Q_{\beta}$), neutron separation energies
($S_{\rm n}$) or $Q_{\alpha}$ values, since they are directly sensitive to the
derivative of  $M_{\rm exp} - M_{\rm calc}$.
These   drastically decreased in the
$Z \simeq 50$, $N \simeq 82$ Sn and in the $Z \simeq 82$, $N \simeq 126$
Pb regions.
For $N > 50$ and $S_{\rm 1n} <5$ MeV the $S_{\rm 1n}$ rms deviation is
0.341 MeV for FRDM(1992), 0.320 MeV for FRDM(2012), 0.443 MeV for HFB21 \citep{goriely09}
and 0.361 MeV for HFB24 \citep{goriely10}.
However, the HFB models are not fully microscopic, because various phenomenological terms are added to the potential energy obtained as HFB solutions. Such terms are the Wigner term with several adjustable parameters, the HFB mass models are surely not microscopic self-consistent models, but rather macroscopic-microscopic just like ours.
How the FRDM improvements ``visibly'' impact  the calculated 
HEW r-process abundances are
discussed later in Section 3.

\subsection{Prediction of $T_{1/2}$ and $P_{\rm n}$ values from \label{2p2}FRDM - QRPA} 

Beta-decay half-lives $T_{1/2}$ and delayed-neutron emission probabilities $P_{\rm n}$ are among the easiest measurable gross $\beta$-decay quantities of neutron-rich nuclei far from
stability. Apart from their longstanding importance for reactor applications, the two properties have also become important in interpreting nuclear-structure features far from stability and in
explosive nucleosynthesis studies. Theoretically, both integral quantities are interrelated via their usual definition in terms of the so-called $\beta$-strength function $[S_{\beta}(E)]$~\citep{duke1970,kratz84}. The half-life  may  yield information on the average $\beta$-feeding of a nucleus. However,
since the transition rates to low-lying single-particle states are strongly
enhanced by the phase-space factor $f \sim (Q_{\beta} - E^{*})^5$, 
where $E^{*}$ is the excitation energy of the daughter final states, 
$T_{1/2}$ is most sensitive to the low-energy region of the
$\beta$-strength function.  The $\beta$-delayed neutron-emission probability 
$P_{\rm n}$ is schematically given by the ratio of the integral
$\beta$-intensity to states above $S_{\rm 1n}$ to the total
$\beta$-intensity.  Again, because of the ($Q_\beta - E^{*})^5$ dependence
of the Fermi function, the $P_{\rm n}$ are mainly sensitive to the
$\beta$-feeding to the energy region just above $S_{\rm 1n}$. However,
taking together the two gross decay properties, $T_{1/2}$ and 
$P_{\rm  n}$, may provide some primary information about the nuclear
structure determining the $\beta$-decay.

Since the first simple, phenomenological approach of the ``Kratz-Herrmann 
Formula'' \citet{kratz73}, with updated compilations by 
\citet{pfeiffer02,mccutchan12}, numerous theoretical 
models have been used to predict $T_{1/2}$ and $P_{\rm n}$ values of unknown 
exotic isotopes; however many of these calculations were limited to  
localized  mass regions. For a detailed discussion of the significance and 
sophistication of such models, see, e.g. \citet{moller03}. 
 In the present study, we use again the FRDM-QRPA formalism as outlined 
in \citet{moller03}, now based on the FRDM(2012) model. 
We again take into account allowed GT and ff transitions, and as previously  
introduce an empirical Gaussian spreading of the individual GT transition 
strengths above 2 MeV, with width roughly corresponding to the model
mass accuracy.
\begin{figure}[t]
 \begin{center}
 \includegraphics[width=\linewidth]{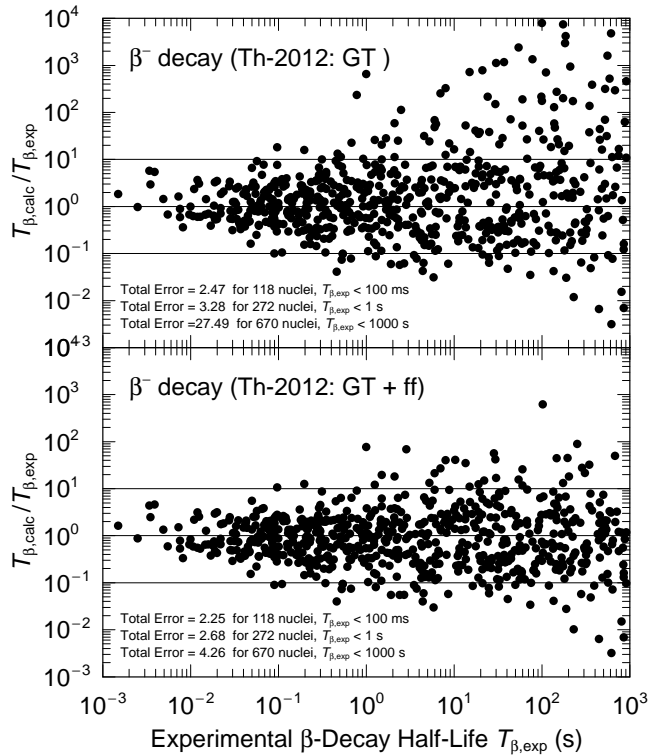}
   \caption{Ratio of calculated to experimental $\beta$-decay half-lives $T_{1/2}$ for nuclei from $^{16}$O to the heaviest known ones. In the upper part, only the theoretical GT-strength is considered, whereas in the lower part GT+ff calculations are used for comparison. The respective total errors are given for three different half-life ranges, i.e. $T_{1/2} < 1000$ s, $< 1$ s and $< 100$ ms, where the last limit represents the usual range for r-process nuclei. For further details, see text and \citep{pfeiffer02,moller03}}.
\label{fig2}
 \end{center}
\end{figure}

In Figure \ref{fig2} we compare measured $\beta$-decay half-lives with calculations based on our models for pure GT and GT+ff transitions, for nuclei throughout the periodic system. To address the reliability versus distance from stability, we present the ratio $T_{\beta,calc}/T_{\beta,\rm exp}$ versus the quantity $T_{\beta,\rm exp}$. As in our earlier work \citep{moller97,pfeiffer02,moller03} we find that the relative deviation between calculated and measured half-lives 
decreases towards smaller  measured T$_{\beta,\rm \rm exp}$, which means the
error decreases with distance from stability. 
Furthermore, our half-life comparisons show that the mean 
deviation of the calculated from the experimental values are approximately 
zero. This indicates, that no ``renormalization'' of the calculated $\beta$-strength
is necessary. The calculated half-lives agree with the 272 experimental
values listed in {\sc Nubase2012}  
with $T_{1/2}\le1$ s to within a factor of 2.68. 
\begin{figure}[t]
 \begin{center}
 \includegraphics[width=\linewidth]{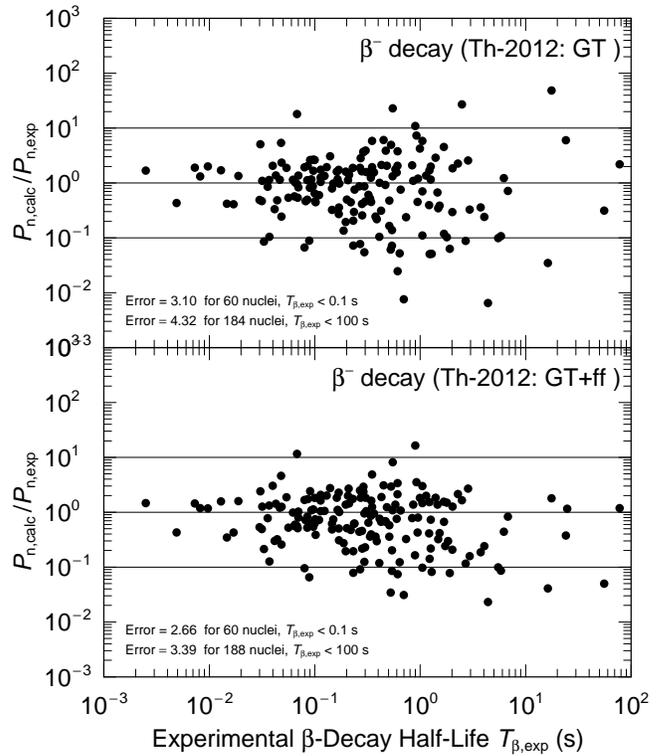}
   \caption{Ratio of calculated to experimental $\beta$-decay neutron-emission probabilities $P_{\rm n}$. In the upper part only the theoretical GT-strength is considered, whereas in the lower part GT+ff calculations are used for comparison. The respective total errors are given for the whole ensemble of $\beta$-decay precursors, and for the limited number of r-process nuclei with $T_{1/2} < 100$ ms. For further details, see text and \citep{pfeiffer02,moller03}}.
\label{fig3}
 \end{center}
\end{figure}

In Figure \ref{fig3}, we compare measured $\beta$-delayed neutron emission probabilities with the calculations based on our two models for nuclei in the fission-fragment region. Again, to address the reliability versus distance from stability we present the ratio $P_{\rm n,calc}/P_{\rm n,\rm exp}$ versus the quantity $T_{\beta,\rm exp}$. Although for the $P_{\rm n}$ quantity, we still have substantially fewer data than for the $T_{1/2}$ values, similar trends are observed when comparing to the predictions for pure GT-decay. When the ff strength is added we again find that the new calculation gives improved agreement with experimental data. The calculated $P_{\rm n}$ values agree with the {\sc Nubase2012} values to within a factor 3.39 for the 188 nuclei, where we in this case compare to the full half-life range.

\subsection{The HEW r-process model}
The basic nucleosynthesis mechanisms for elements beyond Fe by slow
(s-process) and rapid (r-process) captures of neutrons have been known
for long time \citep{burbidge57,cameron57,coryell61}.
However, the search for a robust r-process production site has proven
difficult. Still today, all proposed scenarios not only face problems with
astrophysical conditions,  but also with the necessary
nuclear-physics input for very neutron-rich isotopes. Among the various r-process sites suggested during the past decades, the most recent ones supposedly favored in the nucleosynthesis community, are neutron star mergers (NSM; see, e.g. \citep{korobkin12,goriely2013}) and magnetorotationally driven SNe (see, e.g. \citep{winteler2012}). However, even these scenarios are not fully convincing.
On the one hand, NSMs, apart from the still ongoing debates whether they can be responsible for the r-process observations in metal-poor halo stars in the early Universe, the resulting abundance patterns of the heavier elements differ from the S.S.-r one. This does not exclude the merger scenario; however, it seems unlikely that the NSM r-process represents the dominant origin of the ``main'' r-component with A $\ge$ 130. 
On the other hand, magnetohydrodynamic SNe explosion  models (still) require extremely high magnetic fields and strong rotation to produce bipolar jets with possible r-process matter ejecta. However, also this promising scenario has been questioned, e.g.~\citet{fryer2007} and by the recent 3D simulations of~\citet{moesta2014}.
Therefore, in our r-process parameter studies we prefer to use the neutrino-driven or high-entropy wind (HEW) of core-collapse SNe, which still is one of the best studied ``classical'' mechanisms (for representative publications over the last two decades see, e.g. \citep{woosley94,takahashi94,wanajo01,wanajo06,farouqi09a,farouqi09b,farouqi10}. However, until recently even the most sophisticated hydrodynamical models have predicted that the neutrino-driven wind is proton-rich (electron fraction Y$_e > 0.5$) during its entire life, thus actually precluding a rapid neutron-capture process (see, e.g. \citep{fischer10,roberts10}).
However, recent work on charged-current neutrino interaction rates (see, e.g. \citep{roberts12,pinedo2012}) seems to revive the HEW scenario as a possible r-process site by predicting that $Y_e$ can well reach neutron-rich conditions a few seconds after bounce, with minimal values of $Y_e \simeq 0.42$ but still with too low entropies of $S\le 120$. When considering in addition active-sterile neutrino mixing,~\citet{wu14} even predict $Y_e$ values down to about 0.31. Under such  $Y_{e}-S$ conditions, the production of the light trans-Fe elements from Sr up to about Cd would be possible. However, the heavier (REE and 3$^{\rm rd}$ peak) elements observed in ultra-metal-poor halo stars (see, e.g. \citep{roederer10}) would not (yet) be produced. Furthermore, it is admitted that there are still various model uncertainties about weak rates, nuclear symmetry energy, weak magnetism, inelastic processes, etc.~\citep{fischer2014}. Also the versions of the equation of state (EOS) used in the above calculations do not seem to be the optimum choice, since they  fail several tests related to symmetric nuclear matter, pure neutron matter and/or symmetry energy, and its derivatives \citep{dutra2014}. Therefore, a final conclusion about the realistic $Y_{e}-S-V_{\rm exp}$ parameter space of the cc-SN scenario can presumably not yet be drawn.
Given the present model situation, and since the main goal of the present paper is to compare the results with the old and new FRDM \& QRPA nuclear-physics input to r-process abundance calculations, to us it appears justified to use again our parameterized, dynamical HEW approach, based on the initial model of \citet{freiburghaus99}, which assumes adiabatically expanding homogeneous mass zones with different values of the radiation entropy $S$ (in $k_{\rm B}$/baryon). The code used for the nucleosynthesis calculations (see~\citet{farouqi10}) has been steadily improved, for example to implement experimental data as well as more reliable theoretical
$\beta$-decay properties in certain ``pathological'' mass regions (see, e.g \citet{arndt11}). Furthermore, a better algorithm has been developed for the tracking of the $\beta$-decaying nuclei back to stability with time-intervals small enough to consider very late recaptures of previously emitted $\beta$-delayed neutrons from longer-lived precursors, such as 55s $^{87}$Br or 24s $^{137}$I.

\begin{figure}[t]
 \begin{center}
 \includegraphics[width=\linewidth]{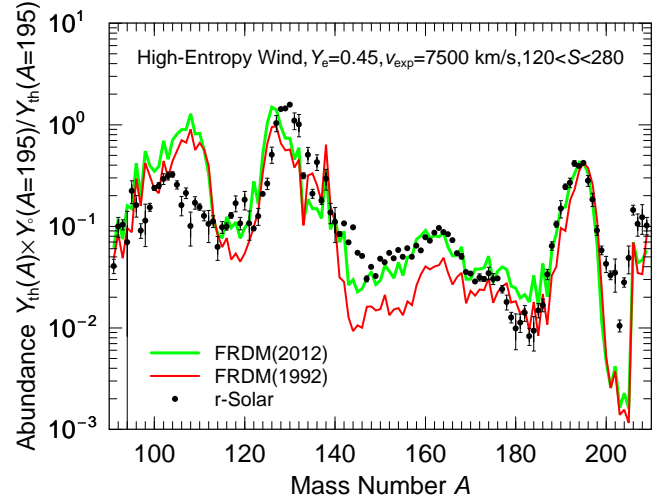}
   \caption{Solar r-process abundances 
compared to calculations based on two nuclear-structure
data bases: our previous FRDM(1992) and the new FRDM(2012).
Both calculated distributions are normalized to the $N_{\rm r,\odot}$
value of $^{195}$Pt at the maximum of the $A \simeq 195$ peak.}
\label{rabund}
 \end{center}
\end{figure}

As has been outlined in detail in \citet{farouqi10}, the overall wind ejecta
represent a superposition of $S$ components, where the main parameters 
$Y_{e}$,
$S$ and the expansion velocity $V_{\rm exp}$ (in km/s) are correlated via the ``r-process
strength formula'' 
$Y_{\rm n}/Y_{\rm r}\propto V_{\rm exp}\times (S/Y_{\rm e})^3$.
Following our traditional attempt to find a possible explanation
of the S.S. isotopic abundance residuals ($N_{\rm r,\odot}\simeq
N_{\odot} - N_{\rm s,\odot}$; \citet{lodders09,bisterzo11}),
 eventually from a weighted superposition of several HEW
conditions and using for the first time the new FRDM-QRPA
nuclear-physics input, we have investigated the whole astrophysics
parameter space as functions of electron abundance 
($0.40 \le Y_{\rm e} \le 0.499$), expansion velocity 
($2000 {\rm km/s} \le V_{\rm exp} \le 30\, 000 {\rm km/s})$.

\section{REPRODUCTION OF THE S.S. R-PROCESS ABUNDANCES}
In this Article, we present first results of our HEW study based on the new
FRDM(2012) and QRPA(2012) nuclear-physics input as outlined in
subsections \ref{2p1} and \ref{2p2}, and our ``standard'' choice of the
astrophysical parameters of $Y_{e} = 0.45$, 
$V_{\rm exp} = 7500$ km/s and $20 \le
S \le 280$ $k_{\rm B}$/baryon. This simulates ``hybrid'' r-process conditions
with freezeout-temperatures (when $Y_{\rm n}/Y_{\rm r} \le 1$) for free
neutrons of $T_9({\rm freeze}) \simeq 0.7$ for the$ A \simeq 130$ abundance
peak.  For this nuclear and astrophysics input, we show in Figure \ref{rabund}
the integrated r-abundances ($N_{\rm r,\rm calc}$) for our new (2012) calculations  together with the
 $N_{{\rm r},\odot}$ values and our results based on the previous FRDM(1992) and
QRPA(2003) nuclear-physics input, both curves normalized to the
 $N_{{\rm r},\odot}$  value of (98\% ``r-only'') $^{195}$Pt.  
Both $N_{\rm r,\rm calc}$ distributions
represent results of the integration 
over the total mass region with a consistent
weighting of equal-sized ejected HEW elements with equidistant
entropies as outlined in
\citet{farouqi10}. Hence, they are not composed of a superposition of
independently fine-tuned fits of local mass regions, which may in each
case yield better agreement with the $N_{{\rm r},\odot}$ pattern, as for
example shown by \citet{arndt11} for the $A\simeq 130$ r-peak or by
\citet{mumpower12} for the REE region alone.  Therefore, at first
sight the new overall $N_{\rm r,\rm calc}$ distribution may not seem
significantly better than the old one, but this impression is
deceiving as will be discussed in the following.

First, let us motivate why in Figure \ref{rabund} 
we only show the r-abundance pattern
for the entropy range $S \ge 120$, i.e. the region of the HEW
components of the ``weak'' and ``main'' neutron-capture components.
As is well known,
in the HEW scenario a significant overproduction of the mass
region below $A \simeq 110$ seems to be unavoidable 
(see, e.g. 
\citet{woosley94,freiburghaus99,farouqi09a,farouqi10}).
This result has commonly been referred to as a model deficiency due
to the $\alpha$-rich freezeout component at low $S$, which -- according
to the definition given in \citet{farouqi09a} -- concerns the
``charged-particle'' and part of the ``weak'' neutron-capture components
with the elements between Sr and Ag. In \citet{farouqi10} it has
been speculated that one could avoid this discrepancy by considering
HEW ejecta as a mixture of correlated $S$ and $Y_{e}$ components.
However, in our present study we are again unable to verify this
assumption. Hence, even with the new nuclear-physics 
(FRDM(2012)--QRPA(2012))
input the local overabundance in the $100\le A \le 110$ mass
region persists. 
Moreover, as is partly indicated in Figure \ref{rabund}, in this mass
region there is also no improvement of the abundance fit relative to
the earlier FRDM(1992)--QRPA(2003) input.

However, in the heavier $A \simeq 120$ mass region where in the past
in practically all r-process calculations with the FRDM(1992) masses,
the large abundance trough occurred (see, e.g. 
\citet{kratz93,wanajo04,farouqi10,winteler2012,nishimura12}),
we now see that with the new FRDM(2012) masses this discrepancy has
practically disappeared. As has in principle already been discussed long
time ago (see, e.g. \citet{kratz93,chen95,pfeiffer97}),
this trough effect is due to subtle details in the predicted behavior
of the decreasing absolute magnitude of the slope $dS_{\rm 1n}(N)/d2N$ 
for odd $N$ of 
the r-process isotopes of Zr 
to Tc between
the strongly deformed $N = 66$ neutron mid-shell and the shape-transition
region towards the next major spherical shell closure at $N = 82$. Taking,
for example the isotopic chain of $_{40}$Zr, with the FRDM(1992) model
predictions the r-process ``boulevard'' with increasing neutron-density
or $S$ components was populated by isotopes up to odd-$N$ $^{113}$Zr$_{73}$
with the predicted lowest $S_{\rm 1n} = 1.36$ MeV and an unusually strong
prolate quadrupole deformation of $\epsilon_2 = 0.36$. Thereafter, instead
of an expected further decrease of the $S_{\rm 1n}$ values, an 
increase of $S_{\rm 1n}$ of
0.78 MeV occurred in FRDM(1992) between $N=73$ and $N=75$,
and $S_{\rm 1n}$ remains large until the $N = 82$ magic shell. The 
consequence of this behavior has been that the Zr r-abundances pile up in
the lower-mass even-$N$ ``waiting-points'' up to $^{112}$Zr$_{72}$; and then the
r-process directly ``jumps'' over 10 mass units to the classical neutron-magic
waiting-point $^{122}$Zr$_{82}$. With the similar behavior of the 
$dS_{\rm 1n}(N)/d2N$ 
slopes
of the neighboring elements in this mass region, in the extreme case of
an instantaneous freezeout no r-progenitor isotope is produced in
the initial r-process ``boulevard'' between $A = 112$ and the $N = 82$ isotones
of Zr to Tc (see, e.g. Figure  \ref{rabund} in 
Chen et al. 1995). The resulting deep
trough occurring under these simplified conditions cannot be completely
filled up by late non-equilibrium captures of free neutrons and/or the
recapture of previously emitted $\beta$-delayed neutrons during a
more realistic treatment of the freezeout phase, as one can see, for
example from Figure  24 in \citet{farouqi10}, and also from 
Figure \ref{rabund} in this Article.

In the FRDM(2012), however, the description of this 
shape-transition region is improved. Taking again the Zr isotopes
as an example, the r-process also runs up to $^{113}$Zr$_{73}$
 with its lowest, but now somewhat higher $S_{\rm 1n} = 1.74$ MeV and a slightly
lower -- but probably still too high -- deformation of $\epsilon_2 = 0.33$
(cf. \citet{pereira09}). Thereafter, again
a local increase of the $S_{\rm 1n}$ occurs, but less pronounced than in
FRDM(1992). Thus, in the Zr chain the next populated r-process
progenitor is $^{119}$Zr$_{79}$. With the similar behavior of
the $S_{\rm 1n}$ slopes of the neighboring elements, in the FRDM(2012)
several r-progenitor isotopes are populating the r-process ``boulevard''
in the $A\simeq 120$ mass region.
As seen in Figure \ref{rabund}, this population 
 together with
a realistic dynamical treatment of the full freezeout phase now
removes the earlier r-abundance trough.

While the occurrence of this trough below the $A\simeq 130$,
$N_{r,\odot}$ peak in earlier approaches could be considered
as a local ``eyesore'' in the calculated r-abundance distribution,
its nuclear-structure correlation to the $N = 82$ shell closure is of
considerable importance for the r-process production of the heavy
elements. It has been shown that the neutron shell strength below
doubly magic $^{132}$Sn strongly influences the formation of this
peak, and acts as an important bottle-neck for the further r-process
matter flow via the rare-earth-element (REE) pygmy-peak region
and the next major peak at $A \simeq 195$ to the actinide r-chronometer region.

In  Figure \ref{rabund} we show the $N_{\rm r,\rm calc}$ pattern 
of the $A \simeq 130$ r-peak for the above ``standard'' astrophysics 
parameter combination, again with both the old and new  
nuclear-physics input. It is immediately evident 
that with the chosen single $Y_{e}$ component, no
significant ``visible'' improvement of the $N_{\rm r,\odot}$ peak is obtained when
compared to the fit with the older FRDM(1992) nuclear-data input. The maximum
of the peak is already reached at $A=126$, instead of $A=130$, and the right
wing of the peak is underproduced. Apart from this remaining deficiency,
however, the change of the overall trend of the $S_{\rm 1n}$ values in 
the $N=82$
region with the new FRDM(2012), and in particular the trend
 of a decreasing shell-gap below doubly- 
magic $^{132}_{\phantom{0}50}$Sn$_{82}$, defined as the difference of the $S_{\rm 2n}$ 
values for $N = 82$ and $N=84$, has the welcome effect of reducing the 
earlier overly strong bottle-neck behavior of the $N = 82$ shell. 
While with FRDM(1992) the shell gap below $Z = 50$ was predicted 
to further increase with decreasing proton number down to $Z = 40$, 
with the new FRDM(2012) the trend is inverted, and is now similar to the 
predictions from the recent HFB versions  and the so-called 
``quenched'' mass formulas, such as ETFSI-Q \citep{pearson96}.

As we have shown already in~\citet{pfeiffer01}, in fact the correlated effects of nuclear masses ($S_{\rm n}$ values) and $\beta$-decay properties ($T_{1/2}$) have important consequences for the further time behavior and 
the amounts of heavier r-abundances in the matter flow out of the 
$A \simeq 130$ peak to the REE region and beyond. 
While the time interval needed from the r-seed composition
of the $\alpha$-rich freezeout to reach the maximum of the 2nd r-peak
still is about the same (for $20 \le S \le 190$, 
$\tau_{\rm r} \simeq 160 $ ms) 
for the old and new nuclear-physics input, the time to
overcome the peak bottle-neck within
the $S$ range $190 \le S \le 210$ is with 
$\tau_{\rm r} \simeq 160$ ms for FRDM(1992) already significantly 
longer than that for FRDM(2012) with only $\tau_{\rm r}\simeq 95$ s. 
This time difference then continues for the formation of the REE 
and the $A \simeq 195$ peaks, resulting in total r-process durations 
(for the whole $S$-ranges of $20 \le S \le 280$) of 
$\tau_{\rm r,tot}  \simeq 680$ ms in the past, 
whereas with FRDM(2012) a ``speeding-up'' of the 
r-process is obtained with a $\tau_{\rm r,tot} \simeq 535$ ms.

With respect to the exact shape and position of the 
$ A \simeq 130$ peak, we can furthermore check why we do not 
meet the maximum of the peak exactly at $A = 130$. A closer look at the $S_{\rm 1n}$ 
values of the $N = 83$ isotopes below $Z = 50$ for FRDM(2012) shows the 
expected (and for the r-process ``favored'') smooth decrease by about 
1.47 MeV between $^{133}_{\phantom{0}50}$Sn and $^{128}_{\phantom{0}45}$Rh. 
However, when continuing further down in $Z$ to $N = 83$ $^{127}$Ru 
and $^{126}$Tc, in FRDM(2012) a sudden drop in $S_{\rm 1n}$ by 0.55 MeV 
occurs, followed by even negative $S_{\rm 1n}$ values for $^{125}$Mo to 
$^{123}$Zr, which is not predicted in ``quenched'' mass formulas.  
Quite obviously, it is this local $S_{\rm 1n}$ behavior that causes
the largest bottle-neck for the r-process at these mass numbers, resulting in 
the observed maximum of the $N_{\rm r,\rm calc}$ peak already at $A = 126$ with 
the FRDM(2012) masses and the above HEW conditions. We know from  earlier r-process studies with the ``quenched'' mass model ETFSI-Q that relative to our standard conditions of a ``hybrid'' r-process variant (with $T_{9,\rm freeze}=0.8$ and a time duration until freezeout of $\tau_{\rm r}=130$ ms) for the condition of a ``hot'' r-process variant (with $T_{9,\rm freeze}=1.2$ and a time duration until freezeout of $\tau_{\rm r}=560$ ms) the maximum of the peak can be shifted from $A=128$ to $A=130$ and even slightly beyond. Therefore, we have also checked the outcome of our HEW calculations with the new FRDM(2012), and have obtained similar results. We thus conclude that within our HEW model the correct shape and position of the $A\simeq 130$ r-abundance peak can only be reproduced by a weighted superposition of different $Y_{e}$-components.

The next subject of our discussion of the $N_{\rm r,\rm calc}$ pattern 
is the pygmy-peak region of the REE in between the two major 
r-peaks. As can be seen from Figure \ref{rabund}, considerably improved agreement 
with $N_{\rm r,\odot}$ is achieved with the new FRDM(2012) 
input. With the FRDM(1992) masses, the whole REE region of 
$140 \le A \le 180$ is underestimated by a mean factor of 1.67, 
whereas with the new nuclear input the overall agreement in shape 
and abundance magnitude is excellent, with a mean factor of only 
0.97. The reason for this improvement clearly lies in the reduced 
bottle-neck behavior of the $A \simeq 130$ major r-process peak, which 
obviously has enabled a higher r-matter transfer to the REE mass 
region within the respective S-range of $210 \le S \le 250$. 
This shows that  earlier speculations that significant feeding from 
fission cycling would be necessary to properly fill up the REE 
region can be excluded.  This assumption 
was introduced to ``repair'' a 
clear nuclear-model  deficiency by addition of 
fission-material originating from unrealistically high $S$ components 
($S \ge 400$) with unusually high weights of r-progenitor isotopes 
in the heavy actinide region. Within the HEW scenario combined with
the FRDM(2012)-QRPA nuclear-structure data the REE peak is well reproduced
without invoking fission recycling.

Both our $N_{\rm r,\rm calc}$ distribution for FRDM(1992) and FRDM(2012) 
have been normalized to the S.S. r-abundance of $^{195}$Pt. 
With this normalization the 
shape and position of the 3rd r-process peak 
are well reproduced, with a slight improvement 
of the rising wing of the peak with FRDM(2012). This 
shows that with FRDM(2012) the shape-transition region before 
the peak and the $N = 126$ shell strength are described correctly.

In contrast, the shape-transition region above $N = 126$ may still be
imperfect, as is evident from the deep r-abundance trough in the
$199 \le A \le 205$  region. 
Just as in the $A \simeq 120$ region, the 
$S_{\rm 1n}$ slopes of the specific elements of $_{68}$Er to  
$_{70}$Yb show a ``bump'' behavior beyond $A \simeq 198$, thus precluding 
significant initial and final r-abundances below the Pb region. In 
addition to the $S_{\rm 1n}$ effect, probably also the onset of 
collectivity beyond the $N = 126$ shell closure seems to set in somewhat 
too early. A possible indication for this speculation is that a HEW 
calculation with FRDM(2012) masses and spherical $\beta$-decay 
properties indicates a significantly better reproduction 
of the $A \simeq 204$ trough region.

As the last point of our comparison of how the $N_{\rm r,\rm calc}$ agrees with the 
$N_{\rm r,\odot}$ pattern, we point out that the predicted underabundances for the Pb and Bi isotopes shown in 
Figure \ref{rabund} occur 
because here the important abundance fraction from the $\alpha$- 
backdecays from the actinides has not (yet) been added.

As a brief summary of our Article we find that 
the new nuclear-structure data base has removed some much-studied differences
between calculated and observed S.S. abundances without resorting to
arbitrary changes of relevant sections in the calculated  data bases
as has been the practice in many previous and present studies. With our use of a fully consistent
model framework, namely the  FRDM(2012) and QRPA(2012) combination as
input for r-process calculations, we again demonstrate that, coupled with detailed 
comparisons with astronomical and cosmochemical observations, this approach is     
a valuable tool for learning about both nuclear structure far from 
stability and the required astrophysical conditions for a full r-process.\\

First, we would like to acknowledge the constructive criticism of the anonymous referee, who greatly helped to improve our paper. Furthermore, we thank Jirina Stone, Tobias Fischer, Matthias Hempel and Mounib El Eid for fruitful discussions. PM  carried out this work under the auspices of the National Nuclear Security Administration of the U. S. Department of Energy at Los Alamos National Laboratory under Contract No.\ DE-AC52-06NA25396.

\end{document}